\documentclass[twocolumn,aps,showpacs,preprintnumbers,amsmath,amssymb,prc]{revtex4-2}

\usepackage{graphicx}
\usepackage{dcolumn}
\usepackage{bm}
\usepackage{natbib}

\begin{document}

\preprint{APS/123-QED}

\title{R-matrix analysis of elastic scattering, phase shift and radiative capture reaction cross sections in the $\alpha + \alpha$ system}

\author{Suprita Chakraborty$^1$}
\email{}
\author{Rajkumar Santra$^2$}
\author{Subinit Roy$^2$}
\email{subinit.roy@gmail.com}
\author{V.M. Datar $^3$}
\affiliation{$^1$B.A.D.V for Girls High, Belgachia, Howrah, India}
\affiliation{$^2$Saha Institute of Nuclear Physics, 1/AF, Bidhan Nagar, Kolkata 700064, INDIA} 
\affiliation{$^3$Institute of Mathematical Sciences, Chennai - 600113, India} 

\date{\today}
     
\begin{abstract}
\noindent 
The unstable nucleus $^8$Be, with its two $\alpha$-cluster configuration, is the doorway to the formation of heavier $\alpha$-cluster nuclei.
Most importantly, its the precursor of the production of $^{12}$C through the Hoyle state, a resonance state of three $\alpha$ clusters, in 
the helium burning phase of a massive star. The nucleus exhibits a ground state band of rotational states established through 
$\alpha-\alpha$ scattering experiments. A subsequent precision particle-$\gamma$ coincidence measurement of the electromagnetic transition 
between the 4$^+\rightarrow$ 2$^+$ excited states also corroborated the evidence for a highly deformed dumb-bell shaped structure of $^8$Be. 
A simultaneous phenomenological R-matrix analysis of the measured capture reaction cross sections along with the elastic excitation function and 
phase shift data has been performed. The resulting reduced transition strength of 21.96$\pm$3.86 $e^2 fm^4$ compares well with the estimated 
experimental value of 21.0$\pm$2.3 e$^2$ fm$^4$. The R-matrix yield of the B($E2$) value is closer to the prediction of cluster model but about 19$\%$ 
less than the {\it ab initio} result.
\end{abstract}

\pacs{26.20.Fj, 24.30.-v, 27.20.+n}
                                                
\maketitle

\section{Introduction}
In the He burning stage of stellar nucleosynthesis \cite {rolfsbook}, the formation of the unstable nucleus ${}^8$Be with two alpha particles, having a life time of about 10$^{-16}$s, is  
the first step to reach the three alpha resonance state, the so called Hoyle state. The latter is the doorway for the production of $^{12}$C nucleus and the subsequent synthesis of heavier elements.  
Apart from its astrophysical implication, the nucleus ${}^8$Be is the simplest $\alpha$-cluster nucleus exhibiting a dumbbell shaped structure. It is also the fundamental building block for the 
alpha cluster structures of heavier self-conjugate $4n$ nuclei \cite{Merchant, beck}.

The $\alpha+\alpha$ system, as evidenced from the scattering studies \cite{Tombrello, Heydenburg, Nilson, Igo}, forms a rotational band based on a 0$^+$ state, a narrow $l$=0 resonance 
with an energy of 92 keV above the $\alpha-\alpha$ threshold \cite{bmv2}. The state corresponds to the ground state of $^8$Be nucleus. Besides the 0$^+$ resonant ground state, the ground state band 
in  ${}^8$Be consists of a broad resonance at 3.03 MeV ($\Gamma$ = 1.51 MeV) and a still broader resonance at 11.4 MeV ($\Gamma$ = 3.5 MeV) \cite{nndc}. The spin parities of the excited 
states of the band have been assigned to J$^{\pi}$=2$^{+}$ and 4$^{+}$, respectively. Phase shift data for the $s-$, $d-$ and $g-$ partial waves of $\alpha-\alpha$ relative motion and the nuclear 
bremsstrahlung cross sections, extracted from the $\alpha-\alpha$ scattering measurements were subjected to various model calculations to obtain the resonance properties of $^8$Be states and 
to probe the nature of the potential between the two $\alpha$-clusters \cite{buck, rae, langanke, bhoi, hirabayashi} .

A precision measurement, using a particle-gamma coincidence technique, of the electromagnetic transition between the very broad 4$^+$ resonance around 11.4 MeV to the broad 2$^+$ resonance 
of the ground state rotational band in $^8$Be was reported by Datar {\it et al.} \cite{datar1, datar2}. The measurement was carried out in the incident alpha energy range of 19 to 29 MeV scanning the radiative
capture channel in the $^4$He+$^4$He reaction around the energy location of the broad 4$^+$ state. The experiments provided the first electromagnetic signature of highly deformed dumbbell shaped 
structure of the states in the ground state band of $^8$Be. Cluster model calculation of Ref.\cite{lang-rolfs} with an $\alpha +\alpha$ potential of \cite{buck} described the measured capture 
cross section nicely in the low energy side but yielded a lower peak cross section and dropped off sharply beyond the resonance peak. Assuming a Breit Wigner form factor for the 4$^+$ resonance, the experimental reduced strength for the 4$^+ \rightarrow$ 2$^+$ $E2$ transition was estimated to be 21 $\pm$ 2.3 e$^2$fm$^4$ corresponding to a gamma partial width of 0.48 $\pm$ 0.05 eV for a 
$\gamma$-branching ratio of 1.37$\times$10$^{-7}$. The $B(E2)$ value obtained from the {\it ab initio} calculation based on the Green's Function Monte Carlo method, presented in the Ref.\cite{datar2}, was 
27.5 $\pm$ 0.15 e$^2$fm$^4$. On the other hand, the cluster model prediction for the same is 18.06 e$^2$fm$^4$ \cite{lang-rolfs}.        
 
In this paper, we present a phenomenological R-matrix analysis of the available radiative capture cross sections of Ref.\cite{datar2}. We have performed a simultaneous analysis of the experimental 
low energy phase shift data as the latter is an important aspect of any meaningful study of nucleus-nucleus resonant scattering process \cite {langanke}. Thus, both
the particle and electromagnetic decay channels of the $\alpha-\alpha$ reaction are fitted within the R-matrix method to constrain the resonance parameters. A description of available 
elastic angular distribution data using the best fit parameters has also been presented. 
 
\section{Analysis}
\begin{figure}
\centering
\includegraphics[width=0.51\textwidth]{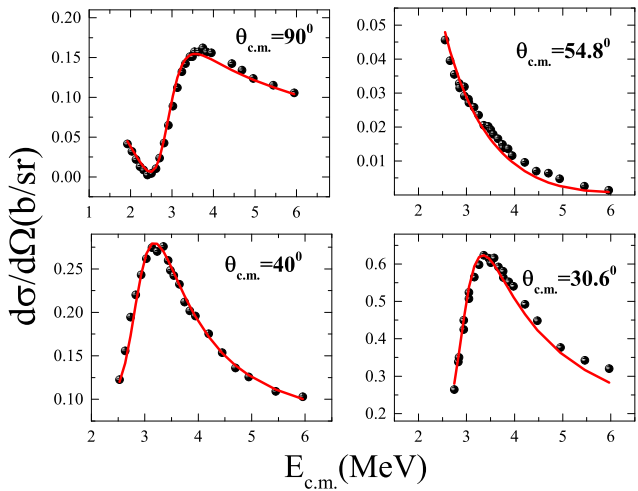}
\caption{\label{fig1} Excitation functions for $\alpha$-$\alpha$ elastic scattering at the centre of mass angles of 90.0$^\circ$, 54.8$^\circ$, 40.0$^\circ$ and 30.6$^\circ$. The solid curves are 
the fits obtained from R-matrix analysis.}. 
\end{figure}
The work presents a simultaneous phenomenological R-matrix analysis of three different categories of data from $\alpha$ on $\alpha$ collisions. The radiative capture cross section data of the reaction 
$\alpha(\alpha, \gamma) {^8}$Be around the location of 4$^+$ resonance at E$_x$ $\sim$ 11.4 MeV, measured through the $\gamma$-ray transition from 4$^+$ to 2$^+$ final state in $^8$Be, are taken from Refs. \cite{datar1, datar2}. The resonant elastic $\alpha-\alpha$ scattering cross sections are from Tombrello {\it et al.} \cite{Tombrello}. The phase shift data for $l$= 0, 2, 4 partial waves from $\alpha-\alpha$ scattering are from Refs. \cite{Tombrello, Heydenburg, Nilson, miyaki, Bacher}. The phase shift data in the energy range between E$_{cm}$ = 0.45 MeV to 2.5 MeV are taken from Heydenburg and Temmer \textit{et al.} \cite{Heydenburg}. The data between E$_{cm}$ = 2.5 MeV to 5.9 MeV are from Tombrello \textit{et al.} \cite{Tombrello} while the data for 6.0 MeV to 11.4 MeV are taken from Nilson \textit{et al} \cite{Nilson}. The higher energy data beyond 11.4 MeV are obtained from Refs. \cite{miyaki, Bacher}. The elastic scattering excitation function data \cite{Tombrello} measured at the centre of mass angles of 30.6$^\circ$, 40.0$^\circ$, 54.8$^\circ$ and 90.0 $^\circ$, respectively, extending from 2 MeV to 6 MeV in energy are also included in the fitting scheme. 

The R-matrix theory has been developed in the seminal works of Lane and Thomas \cite{lane}, Azuma {\it et al.} \cite{Azuma} and Descouvemont and Baye \cite{descouvemont} and further elucidated in a
recent review work by de Boer {\it et al.} \cite{deboer}. While avoiding a detailed discussion on R-matrix formalism, it should be mentioned that we have adopted the formalism given 
in Azuma {\it et al.} and used the multilevel, multichannel R-matrix code AZURE2 \cite{Azuma} for the phenomenological analysis.  

In phenomenological analyses of low energy nuclear reaction data, R-matrix theory is used by adjusting the parameters of the model, the pole energies and the reduced widths of the decay channels, 
to optimize the agreement with the experimental data. Some of the recent applications highlighting the procedure are found in Refs. \cite {deboer, chakraborty15, li16, chakraborty2}. Two important 
aspects of the phenomenological R-matrix analysis are the choices of channel radius and the background poles. These are correlated quantities due to their origins in the R-matrix model. Therefore, 
a sensitivity check should be performed for phenomenological R-matrix fit to the data \cite{deboer}. 

The {\it channel radius} basically divides the nuclear configuration space into two distinct regions \cite{lane} - the internal region bound by the channel radius where nuclear interaction is dominating and nuclear many body effects are important and the external region beyond the channel radius where the nuclear interaction either vanishes or is extremely insignificant and Coulomb interaction is the predominant interaction. A matching of the interior wave function with the exterior wave function at the channel radius describes the R-matrix. It represents the information on the structure of the compound nucleus. In R-matrix theory, the representation of a physical state of the compound nucleus with a definite $J^{\pi}$ value in terms of a complete orthonormal set of R-matrix states involves a sum of a large number of terms. Truncation of this very large number of virtual states in the expansion requires the introduction of {\it background states or poles} \cite{lane}. Identifying a finite number of low energy physical states with the lowest R-matrix states, the effect of high lying levels in the truncated scenario is taken into account through one or more broad high energy virtual states or background poles \cite {deboer}. A background pole is introduced for each $J^{\pi}$ set of physical states. The parameters of the phenomenological model of R-matrix formalism can be related to the observable quantities. In the present analysis with AZURE2, the alternative approach of Brune \cite{brune}, where all the parameters of the model have simple physical meaning, has been adopted for the required parameter transformation.

To describe the data, altogether eight resonances in $^8$Be have been considered. Besides the 0$^+$, 2$^+$ and 4$^+$ states of the ground state band, one 0$^+$ resonance at 20.0 MeV excitation, three more 2$^+$ resonances in the excitation energy window of 16 MeV to 23 MeV and  a 4$^+$ resonance at 19.2 MeV \cite {nndc} have been included. Energies (E$_x$) of the first 2$^+$ and 4$^+$ states and particle decay widths ($\Gamma_p$) of the resonant 0$^+$ ground state, first 2$^+$ and 4$^+$ resonance states are varied to obtain the best fits. The partial $\gamma$-decay width of the 4$^+ \rightarrow$2$^+$ transition is also varied to fit the capture reaction data but the decay width for the 2$^+ \rightarrow$0$^+$ $\gamma$-transition in the band is kept fixed at the estimated value of 8.3 meV \cite{lang1} because no experimental data exists for the transition. The energy location and particle decay width  of the second 2$^+$ state near 16 MeV have been kept fixed during the search. It is to be pointed out here that the present analysis required the presence of only one 2$^+$ resonance to describe the $l$=2 phase shift data around 16 MeV, unlike those of 16.626 MeV and 16.922 MeV in Ref. \cite {tilley}.  While both the energies and particle decay widths of the second 4$^+$ state near 19 MeV and the fourth 2$^+$ state around 22 MeV have been varied, only the particle decay widths of the second 0$^+$ and third 2$^+$ states around 20 MeV excitation are searched during the fitting procedure. The values of the parameters kept fixed during the search are taken from Ref. \cite{nndc}. The parameters that are varied to obtain the best simultaneous fits to the data sets are marked in bold in Table \ref{tab1}.

A proper choice of radius for entrance channel is needed for the R-matrix model calculation. However, the channel radius is not a free parameter of the model. According to the formal R-matrix theory channel radius should be large enough making the nuclear force negligible beyond that. A large channel radius may make the required number of background poles overly large, as the two are correlated, to cancel out most of the effect of hard sphere scattering phase shift component \cite{deboer}. To fix the radius of the $\alpha+\alpha$ channel, we followed a procedure mentioned in Ref.\cite{deboer}, which achieves a good R-matrix fit that is insensitive to the channel radius. In our analysis, we have introduced four background poles corresponding to the spin parities of J$^\pi$ = 0$^+$, 2$^+$ and 4$^+$, as shown in Table \ref{tab1}. It is to be noted that the locations of the background poles are kept fixed at excitation energies beyond the highest energy real resonance state considered for a particular J$^\pi$ value. Also, the contributions of the background poles in the capture channel have been neglected as their influence on the capture data has been tested to be insignificant. We, then performed a grid search on the channel radius by changing the value in small steps and varying all the search parameters to get the fits to the data sets simultaneously. The particle widths of the background poles, important for the scattering partition, are also varied as free parameters. It is observed that by changing the widths of the background poles with changing channel radius, the total $\chi^2$ value reaches almost a plateau in a range of radius value of 3.40 fm to 4.00 fm. The fits become more or less insensitive to the change in the radius within this range. Considering the overall quality of the fits for these radius values, a value of 3.73 fm is chosen as the radius for the entrance channel that also gives the correct fall off of the phase shift data at higher energies. We would also like to emphasize that two background poles for J$^\pi$=2$^+$ have been used in the present analysis to account for the effect due to the presence of closely spaced resonance structures in the high energy domain of the L=2 phase shift.   

\begin{figure}{c}
\includegraphics[width=0.52\textwidth]{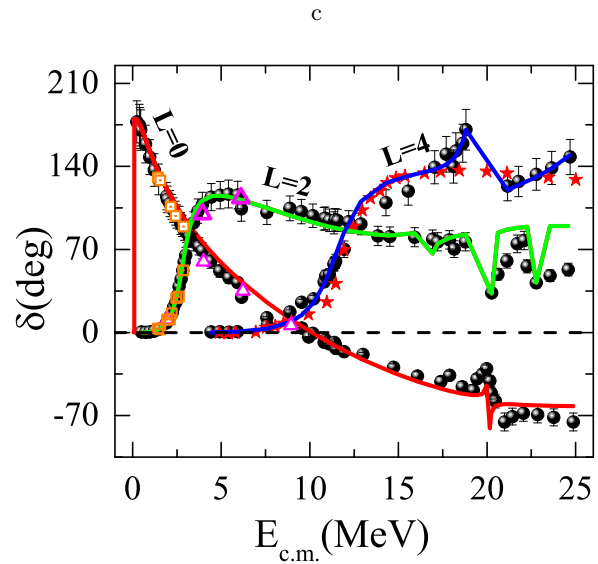}
\caption{\label{fig2} Fits (solid curves) to the phase shift data (black bullets) from Refs. \cite {Tombrello, Heydenburg, Nilson}, and \cite {miyaki, Bacher} (see text for the details) of $\alpha$-$\alpha$ scattering for the partial waves $l$= 0, 2 and 4 in the energy range of 0.45 to 25 MeV. The low energy data points of Russell, {\it et al.} \cite{russell} (orange dotted square) and Jones {\it et al.} \cite{jones} (magenta triangle) for $s$- and $d$- wave phase shifts, respectively are plotted on the calculated curves for comparison only . Pink stars on the $g$-wave phase shift curve are the calculated points for $l$=4 phase shift from \cite{bhoi}.}  
\end{figure}    
\begin{table*}
\caption{The best fit resonance parameters obtained from the present $R$ matrix analysis. The parameters varied during the search for the best fits are shown in bold.}
\label{tab1}       
\vspace*{0.1cm}
\begin{tabular}{c|cccc|cccc}\hline \hline
&&&&&&&& \\
 $J^{\pi}$&\multicolumn{4}{c|}{Present Work}&\multicolumn{4}{c}{From Refs.\cite{nndc, tilley, standard value}} \\  \hline
 &E$_{x}$&$\Gamma_{\alpha}$&$\Gamma_{\gamma} (R\rightarrow0.0)$&$\Gamma_{\gamma}(R\rightarrow3.04)$&E$_{x}$&$\Gamma_{\alpha}$&$\Gamma_{\gamma}(R\rightarrow0.0)$&$\Gamma_{\gamma}(R\rightarrow3.04)$ \\
&(MeV)&(MeV)&(eV)&(eV)&(MeV)&(MeV)&(eV)&(eV) \\ \hline
&&&&&&&&\\
0$^{+}$&0.0&{\bf (4.665$\pm$0.006)}{\bf $\times 10^{-6}$}&&&0.00&5.57$\times$10$^{-6}$&& \\
&&&&&&&& \\
&20.0&{\bf (65.0$_{\bf -12.5}^{\bf +10.4}$)}${\bf \times10^{-3}}$&&&&&&\\
&&&&&&&&\\
\cline{1-5}
&&&&&&&&\\
2$^{+}$&{\bf 3.04$\pm$0.001}&{\bf 1.569$\pm$0.004}&0.0083\footnote{Calculated value of the $\gamma$-width corresponding to {\it E2 } strength of 75 W.U. taken from Ref.\cite{lang1}}&&3.03&1.5&& \\
&&&&&&&& \\
&16.63&108.0$\times 10^{-3}$&&&&&&\\
&&&&&&&&\\
&20.10&{\bf (54.53$\pm$1.60)}${\bf \times 10^{-3}}$&&&&&&\\
&&&&&&&&\\
&{\bf 22.68$\pm$1.02}&{\bf (18.88$_{\bf -0.46}^{\bf +0.49}$)}${\bf \times 10^{-3}}$&&&&&&\\
&&&&&&&&\\
\cline{1-5}
&&&&&&&&\\
4$^{+}$&{\bf 11.62$\pm$0.04}&{\bf 2.67}${\bf \pm}${\bf 0.04}&&\bf {0.44}$_{\bf -0.06}^{\bf +0.07}$&11.4&3.50&&0.48 \\
&&&&&&&& \\ 
&{\bf 19.86$_{\bf -0.06}^{\bf +0.05}$}&{\bf (757.82$\pm$46.3)}${\bf \times 10^{-3}}$&&&&&& \\
&&&&&&&& \\ 
\cline{1-5}
&&&&&&&& \\
&\multicolumn{4}{c|}{Background poles in R-matrix fit}& \\
&&&&&&&&\\
\cline{1-5}
&&&&&&&&\\
0$^{+}$&30.0&{\bf 22.0}&&&&&& \\
&&&&&&&& \\
2$^{+}$&28.0&{\bf 21.0}&&&&&& \\
&&&&&&&& \\
2$^{+}$&30.0&{\bf 15.0}&&&&&& \\
&&&&&&&& \\
4$^{+}$&30.0&{\bf 9.02}&&&&&& \\ 
&&&&&&&& \\ \hline
\end{tabular}
\end{table*}
AZURE2 uses the parameter optimization routine MINUIT2 to perform the least square minimization to generate the best fit to the data from the theoretical R-matrix calculations \cite{azure-man}. MINUIT2 \cite{james} has built in methods for parameter uncertainty calculation. The uncertainties quoted with the parameters in Table \ref{tab1} are from the output of the code. 
\begin{figure}
\includegraphics[width=0.52\textwidth]{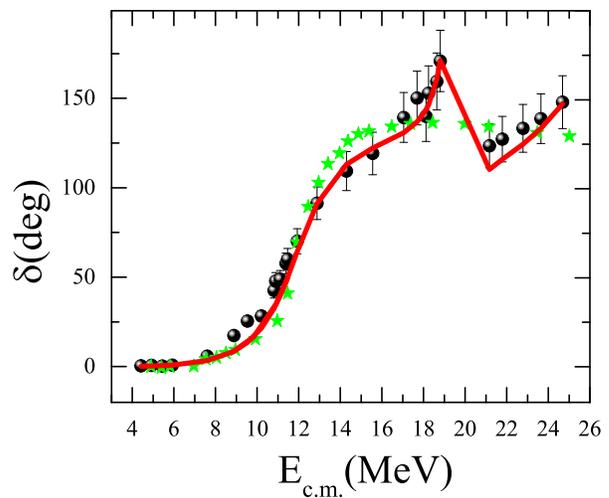}
\caption{\label{fig3} R-matrix description (solid curve) of L=4 phase shift data.} 
\end{figure} 

\section{Results and Discussion}

\begin{figure}
\includegraphics[width=0.5\textwidth]{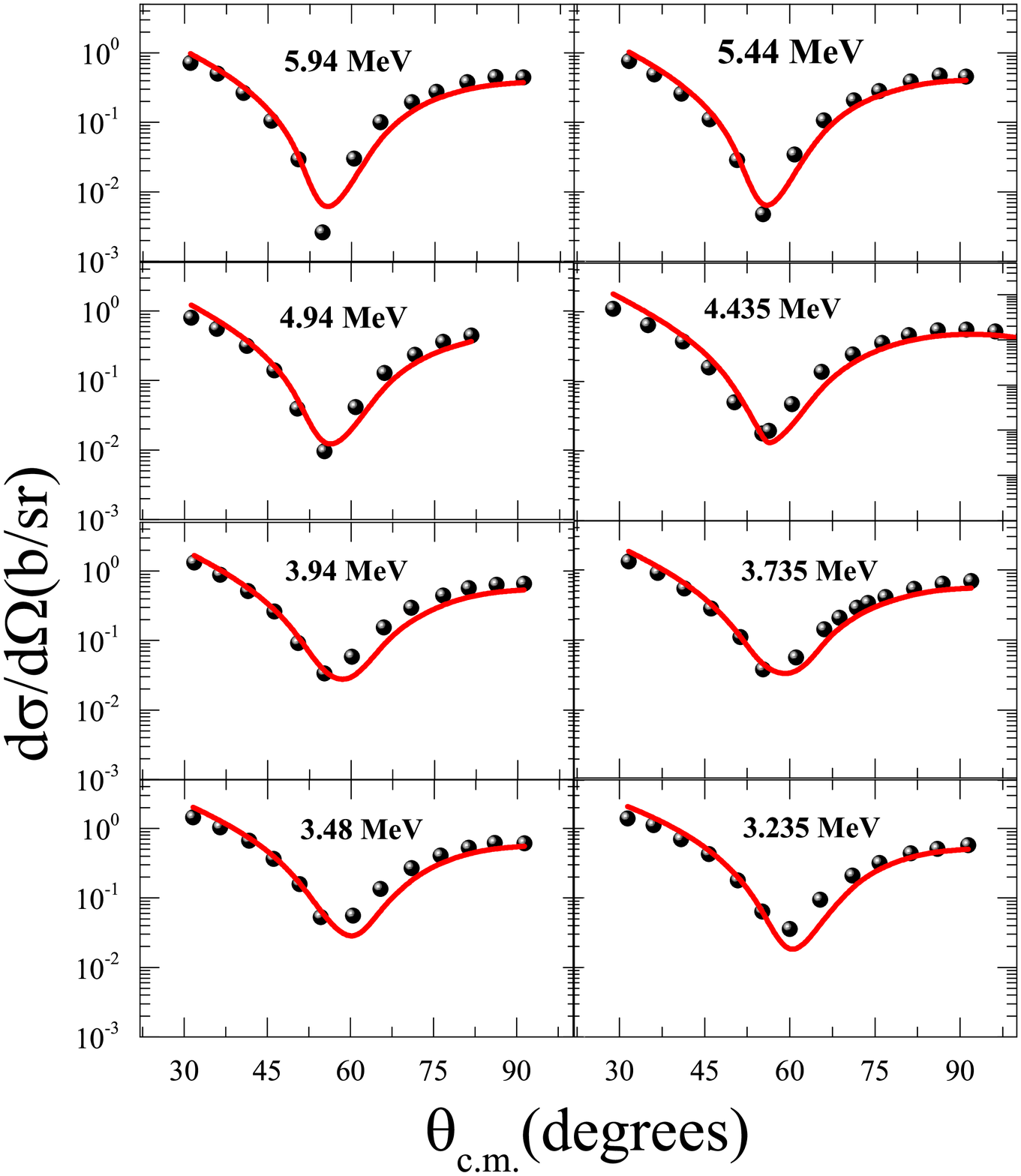}
\caption{\label{fig4} R-matrix description (solid curve) of elastic angular distribution data from Ref.\cite{Tombrello}.} 
\end{figure}
The resultant parameters yielded by the best fits to different sets of data are presented in Table \ref{tab1}. The parameters in bold are, as mentioned earlier, the fit parameters. Others are kept fixed during the 
search for the best fits. The adopted values in the literature for the resonant ground state, the first 2$^+$ and  4$^+$ states of $^8$Be are also shown in the table. The high energy poles are found to be 
necessary to generate the fits to the higher energy phase shift data and in the process optimize the resonance energies and the particle widths of the states in the ground state band of $^8$Be. It is observed 
that the choice of the partial decay widths depend significantly on the choice of the channel radius. Also, this dependence is more sensitive to the particle decay widths rather than the $\gamma$-decay widths of these high energy resonances. Hence, different combinations of channel radius and particle decay widths of the high energy poles including the background have been tested \cite{deboer}. The particle decay widths quoted for the background poles are the best fit values corresponding to a channel radius of 3.73 fm.       

The fits to the elastic excitation function data at the centre of mass angles $\theta_{cm}$= 90$^{\circ}$, 54.8$^{\circ}$, 40$^{\circ}$ and 30.6$^{\circ}$, respectively in the energy range of E$_{c.m.}$= 2.0 MeV to 6.0 MeV, are shown in Fig.\ref{fig1}. In Fig.\ref{fig2}, the resultant fits to the phase shift data have been displayed. The calculated $s$-wave phase shift shows the usual fall off following the data beyond 92 keV. It is positive at lower energies and becomes negative around E$_{c.m.}$ = 10 MeV, observed as well in Refs.\cite{Tombrello, bhoi, hirabayashi}. Introduction of the second ${0^+}$ state at 20.0 MeV in the fitting scheme reproduced the observed structure in the high energy $l$=0 phase shift data. Excellent reproduction of the $d$-wave phase shift is obtained when the three higher energy $2^+$ resonances, taken from \cite{nndc}, are included along with the lowest energy $2^+$ resonance at 3.04 MeV. Although in the higher energy region the model phase shift reproduced the locations of the resonances reasonably well, it could not generate the exact behaviour of the data points around these resonances. In Fig. \ref{fig2}, the low energy data points of Russell, {\it et al.} \cite{russell} (orange dotted square) and Jones {\it et al.} \cite{jones} (magenta triangle) for $s$- and $d$- wave phase shifts, respectively, have been shown for comparison with the calculated phase shifts. Reasonably good overlaps have been obtained. 

The simultaneous R-matrix calculation also fitted the $g$-wave phase shift data, extending up to  E$_{c.m.}$ = 25 MeV, quite nicely except the data points in the 12 to 17 MeV energy window. We would like to highlight that an exclusive phenomenological R-matrix fit to the phase shift data of partial wave $l$=4 has been performed and the fit is shown in Fig.\ref{fig3}. In order to fit the data points in this energy window, which constitutes the rising part around the first 4$^+$ resonance in $^{8}$Be, the resonance energy of the first 4$^+$ state shifts to 12.2 MeV. This will relocate the peak position in the capture cross section curve to higher energy with a deteriorating fall off beyond the peak value. Although Miyaki, {\it et al.} reported an energy level of 12.5 MeV for the state (with a hard sphere radius of 3.5 fm), the present simultaneous analysis of scattering and capture data does not corroborate the result of Ref.\cite{miyaki}. Hence, in the simultaneous analysis, we sacrificed the quality of the fit to the $g$-wave phase shift in this energy domain. The profile of the $g$-wave phase shift data around the resonance at 19.86 MeV with spin parity 4$^+$ \cite{nndc} has been very well described by the model. The $l$=4 phase shift plot obtained from the present R-matrix calculation has also been compared with the potential model prediction for $g$-wave phase shift from Ref. \cite{bhoi} (pink star) in Fig.\ref{fig2}. Reasonable matching has been obtained up to around E$_{c.m.}$ = 17 MeV, but the potential model calculation could not generate the resonant structure around 19 MeV. 

The goodness of the resonance energy locations and the associated particle decay widths obtained from the R-matrix fitting have been tested through the reproduction of the elastic angular distributions from Ref. \cite{Tombrello}, shown in Fig.\ref{fig4}. Quite good overall description of the angular distribution data is observed except a tendency of shift in the positions of the minima towards the higher angles as the energy decreases.
\begin{figure}         
\includegraphics[width=0.5\textwidth]{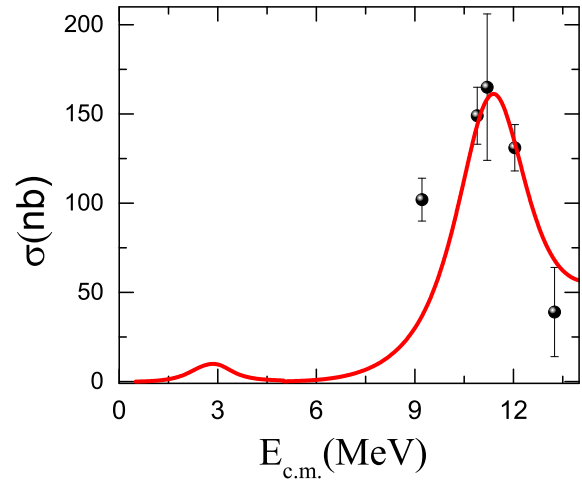}
\caption{\label{fig5} R-matrix fit to 4$^+ \rightarrow$ 2$^+$ capture reaction cross section data of $\alpha$($\alpha,\gamma$)$^8$Be from Ref.\cite{datar2}. The calculation is extended to lower energies to
include the 2$^+ \rightarrow$ 0$^+$ capture reaction cross section.} 
\end{figure}

Finally, the fit to the $E2$ capture cross sections for $\alpha({\alpha, \gamma}){^8}$Be reaction as a function of incident $\alpha$-energy in the range of E$_{c.m.}$ = 9.0 to 13.5 MeV around the 4$^+$ resonance state in $^8$Be is shown in Fig.\ref{fig5}. While the cross section curve from the simultaneous R-matrix calculation describes the maximum value of the cross section data well, the lower energy rise of the cross section in the resonance region appears to under predict the data. The fall off of the model cross section in the higher energy region describes the data well. The shape of the excitation curve of R-matrix model cross section in the 4$^+$ resonance region, with a particle width of $\Gamma_p$=2.669$\pm$0.036 MeV, is some what narrower than that predicted by the cluster model (see Fig. 4 of Ref.\cite{datar2}). The best fit value of the $\gamma$-partial width is $\Gamma_\gamma$ = 0.502$\pm$ 0.084 eV, which is closer to the experimentally estimated width of 0.48$\pm$0.05 eV, which is based on a simplistic Breit-Wigner resonance, and leads to a branching ratio $\frac{\Gamma_\gamma} {\Gamma}$ = 1.88$\pm$0.32$\times$10$^{-7}$. The value is slightly on the higher side compared to the estimate of 1.37 $\times$ 10$^{-7}$ obtained from the data assuming a Breit Wigner shape for the 4$^+$ resonance \cite{datar2}. The reduced $E2$ transition strength or B($E2$) value from the R-matrix analysis is 21.96$\pm$3.86 e$^2$ fm$^4$, which is within the error bars of the experimental value of 21.0 $\pm$ 2.3 e$^2$fm$^4$. But the B($E2$) value from the R-matrix analysis is relatively smaller than the value of 27.0 e$^2$fm$^4$ predicted by the {\it ab initio} calculation presented in Ref.\cite{datar2}, while the value is comparable with the cluster model prediction \cite{lang-rolfs}. The excitation energy curve is then extrapolated to the low energy region encompassing the energy region associated with the $E2$ transition from the resonance state around the excitation of E$_{x}$ $\sim$ 3.0 MeV to the ground state of $^8$Be. The peak cross section for the $2^+$ resonance is found to be 10.3 nb, slightly lower than the value of 14 nb, estimated in Ref. \cite{lang1} near 2.7 MeV. Adding the two contributions of the $E2$ transitions corresponding to 4$^+ \rightarrow $2$^+$ and 2$^+ \rightarrow $0$^+$ using the parameters of Table \ref{tab1}, the total excitation curve for $E2$ capture has been generated and shown in Fig.~\ref{fig5}. 

In conclusion, a simultaneous phenomenological R-matrix analysis has been performed with the excitation functions and the phase shift data from the elastic scattering of $\alpha-\alpha$ system and the radiative capture cross sections around the 4$^+$ resonance state of the unbound $^8$Be nucleus from $\alpha(\alpha, \gamma){^8}$Be reaction. A consistent set of parameters is obtained from the simultaneous analysis, however, the resultant $\alpha$-widths of 0$^+$ ground state and the 4$^+$ excited state are lower than the generally accepted values. The $\gamma$-width of the 4$^+ \rightarrow$2$^+$ and hence the reduced transition strength values are comparable with experimental estimates highlighting the largely deformed structure of the excited state of $^8$Be nucleus. The value of the reduced transition strength is closer to the experimental value but lower than the prediction of an {\it ab initio} calculation. An independent R-matrix fit to the $g$-wave phase shift data alone predicted a location for the first 4$^+$ resonance at 12.2 MeV with an indication of larger width. The profile of the state does not match the results from the available capture data and also the present simultaneous analyses. More number of capture cross section data around this resonance can resolve the issue. Also a more consistent R-matrix analysis can be performed if the radiative capture cross section data are available  around the first 2$^+$ resonance in $^8$Be nucleus. This is definitely an extremely challenging experiment, requiring a very high current alpha beam, but worth attempting! It will also probe the predictions of different models for the reduced transition probability of 2$^+ \rightarrow$ 0$^+$ ground state.

 \section*{ACKNOWLEDGEMENT}
The authors would like to thank Richard deBoer, Department of Physics, University of Notre dame, USA for the suggestions and support provided during the $R$-matrix 
calculation using {\it AZURE2}. One of the authors, S.C., would also like to acknowledge the support extended by Saha Institute of Nuclear Physics, Kolkata during the period of work.

\end{document}